# Diluted II-VI Oxide Semiconductors with Multiple Band Gaps

K. M. Yu,[1] W. Walukiewicz,[1] J. Wu,[1] W. Shan,[1] J. W. Beeman,[1] M. A. Scarpulla,[1,2] O. D. Dubon,[1,2] and P. Becla[3]

[1]Materials Sciences Division, Lawrence Berkeley National Laboratory, Berkeley, California 94720
[2]Department of Materials Science and Engineering, University of California, Berkeley, California 94720
[3]Department of Materials Science and Engineering, Massachusetts Institute of Technology, Cambridge, Massachusetts 02139

Abstract

We report the realization of a new multi-band-gap semiconductor. The highly mismatched alloy $Zn_{1-y}Mn_yO_xTe_{1-x}$ has been synthesized using the combination of oxygen ion implantation and pulsed laser melting. Incorporation of small quantities of isovalent oxygen leads to the formation of a narrow, oxygen-derived band of extended states located within the band gap of the $Zn_{1-y}Mn_y$Te host. When only 1.3% of Te atoms is replaced with oxygen in a $Zn_{0.88}Mn_{0.12}$Te crystal (with band gap of 2.32 eV) the resulting band structure consists of *two direct band gaps* with interband transitions at ~1.77 eV and 2.7 eV. This remarkable modification of the band structure is well described by the band anticrossing model in which the interactions between the oxygen-derived band and the conduction band are considered. With multiple band gaps that fall within the solar energy spectrum, $Zn_{1-y}Mn_yO_xTe_{1-x}$ is a material perfectly satisfying the conditions for single-junction photovoltaics with the potential for power conversion efficiencies surpassing 50%.

Recently a new class of semiconductors has emerged, whose fundamental properties are dramatically modified through the substitution of a relatively small fraction of host atoms with an element of very different electronegativity, the so called highly mismatched alloys (HMAs). III-V and II-VI alloys in which group V and VI anions are replaced with the isovalent N [1] and O [2], respectively are the well known examples of the HMAs. For example, $GaN_xAs_{1-x}$ exhibits a strong reduction of the band gap by 180 meV when only 1% of the As atoms is replaced by N [1].

The unusual properties of HMAs are well explained by the recently developed band anticrossing (BAC) model [3-5]. The model has also predicted several new effects that were later confirmed by experiments [6-8]. According to this model the electronic structure of the HMAs is determined by the interaction between localized states associated with N or O atoms and the extended states of the host semiconductor matrix. As a result the conduction band splits into two subbands with distinctly non-parabolic dispersion relations [3].

In most instances, e.g. N in GaAs or O in CdTe, the localized states are located within the conduction band and the anticrossing interaction results in the formation of a relatively wide lower subband [5]. The subband is shifted to lower energies leading to a reduction of the energy band gap. The BAC model predicts that a narrow band can be formed only if the localized states occur well below the conduction band edge. Such a case is realized in ZnTe, MnTe and $Zn_{1-y}Mn_yTe$ alloys where the O level is located roughly 0.2 eV below the conduction band edge.

We have shown recently that pulsed laser melting (PLM) followed by rapid thermal annealing (RTA) is well suited for the synthesis of HMAs. The combined ion beam and laser processing approach has been demonstrated as an effective approach to synthesize dilute semiconductor alloys including $GaN_xAs_{1-x}$ [9,10] and $Ga_{1-x}Mn_xAs$ [11]. Large enhancement by a factor of five in the incorporation of N in $N^+$-implanted GaAs was observed. This is attributed to the rapid recrystallization rate associated with this process which results in the incorporation of impurity atoms to a level well above the solubility limit [12,13].

Here we report the design and synthesis of a new type of material, the highly mismatched $Zn_{1-y}Mn_yO_xTe_{1-x}$ that has a narrow band of extended states within a semiconductor band gap. This material satisfies the criteria for a multi-band semiconductor and could be used to test the theoretical predictions of enhanced solar cell efficiency. The design of our material is based on the band anticrossing (BAC) model of highly mismatched semiconductor alloys (HMAs).

We used O ion implantation followed by pulsed laser melting to synthesize $Zn_{1-y}Mn_yO_xTe_{1-x}$ alloys. Multiple energy implantation using 90 and 30 keV $O^+$ was carried out into $Zn_{1-y}Mn_yTe$ (y=0 and 0.12) single crystals to form ~0.2 μm thick layers with relatively constant O concentrations corresponding to O mole fractions of 0.0165-0.044. The $O^+$-implanted samples were pulsed-laser melted in air using a KrF laser (λ= 248 nm) with a FWHM pulse duration ~38ns [10]. After passing through a multi-prism homogenizer, the fluence at the sample ranged between 0.020 and 0.3 J/cm$^2$. Some of the samples underwent RTA after the PLM at temperatures between 300 and 700°C for 10 seconds in flowing $N_2$.

The band gap of the films was measured at room temperature using photomodulated reflectance (PR). Radiation from a 300 Watt halogen tungsten lamp dispersed by a 0.5 m monochromator was focused on the samples as a probe beam. A chopped HeCd laser beam ($\lambda$=442 or 325 nm) provided the photomodulation. PR signals were detected by a Si photodiode using a phase-sensitive lock-in amplification system. The values of the band gap and the line width were determined by fitting the PR spectra to the Aspnes third-derivative functional form [14].

Fig. 1 shows a series of PR spectra from $Zn_{0.88}Mn_{0.12}Te$ samples implanted with 3.3% of $O^+$ followed by PLM with increasing laser energy fluence from 0.04 to 0.3 $J/cm^2$. Two optical transitions at ~1.8 and 2.6 eV can be clearly observed from the samples after PLM with fluences $\geq 0.08 J/cm^2$. These transitions occur at energies distinctly different from the fundamental band gap transition ($E_M$=2.32 eV) of the matrix $Zn_{0.88}Mn_{0.12}Te$. Identical PLM treatments on unimplanted and $Ne^+$- implanted ZnMnTe samples do not show such transitions indicating that they are not caused by the implantation damage but can be attributed to the presence of oxygen. The results in Fig. 1 suggest that $Zn_{0.88}Mn_{0.12}O_xTe_{1-x}$ layers are formed after $O^+$-implantation and PLM with energy fluence $\geq 0.08 J/cm^2$. The two optical transitions can be attributed to transitions from the valence band to the two conduction subbands, $E_+$ (~2.6 eV) and $E_-$ (~1.8 eV) formed as a result of the hybridization of the localized O states and the extended conduction band states of ZnMnTe. We note that the observation of the strong photomodulated transition signals at both $E_-$ and $E_+$ indicates the extended nature of these electronic states and the substantial oscillator strength for the transitions. No optical transition is observed for the implanted samples with PLM fluence lower than $0.04 J/cm^2$ indicating that the melting threshold for ZnMnTe is between 0.04 and $0.08 J/cm^2$.

The observed optical transitions shown in Fig. 1 are fully consistent with the predictions of the BAC model for the ZnMnOTe alloys. Although both the $E_-$ and $E_+$ transitions have been observed for III-N-V HMAs [3], only the $E_-$ transition has been reported in II-VI HMAs [2,4]. The transitions at around 2.6 eV shown in Fig. 1 are the first observation of the $E_+$ subband in II-VI HMAs.

The BAC model suggests that the two level interaction leads to the upward shift of the upper state and the downward shift of the lower state by exactly the same energy equal to $\frac{1}{2}\left[\sqrt{(E_a - E_b)^2 + 4C_{LM}^2 x} - |E_a - E_b|\right]$, where $E_a$ and $E_b$ are the original energy levels (in this case $E_M$ and $E_O$), and $C_{LM}$ is the matrix element describing the coupling between localized states and the extended states [3-5]. PL studies in ZnTe have shown that at doping concentrations, oxygen forms a localized level in the gap with an energy $E_O$ =2.0 eV above the VB edge [15]. From the known band offset of ZnTe and MnTe the O level is estimated to be at 2.06 eV for ZnTe with 12% of Mn. The PR spectrum from the $O^+$-implanted ZnMnTe sample PLM with 0.15 J/cm$^2$ shown in Fig. 1 indicates very similar energy shifts ($E_O$-$E_-$)=0.234 eV and ($E_+$-$E_M$)=0.24 eV. This is in perfect agreement with the prediction of the BAC model.

For GaN$_x$As$_{1-x}$ and ZnSe$_x$Te$_{1-x}$ the values of the coupling parameter $C_{LM}$ have been determined to be 2.7 and 1 eV, respectively [3,4]. In this work $C_{LM}$ cannot be determined independently because a precise measurement of the fraction of O on the Te sublattice (i.e., $x$) is not possible. Since it is believed that the magnitude of $C_{LM}$ depends on the electronegativity difference ($\Delta\chi$) between the matrix anion elements and this difference is larger between O and Te ($\Delta\chi$=1.4) than that between N and As ($\Delta\chi$=1.0), we make the assumption that $C_{LM}\approx$3.5eV in ZnMnTe. The substitutional O content of the $Zn_{0.88}Mn_{0.12}O_xTe_{1-x}$ alloys formed by $O^+$-implantation followed by PLM with 0.15 J/cm$^2$ shown in Fig. 1 is estimated to be x≈0.01.

We have reported previously that the formation of $Cd_{1-y}Mn_yO_xTe_{1-x}$ alloys by $O^+$-implantation and RTA, Mn leads to increased stability of substitutional O in $Cd_{1-y}Mn_yTe$ [2]. In the present work we have also demonstrated the formation of $ZnO_xTe_{1-x}$ layers with x~0.01 by $O^+$-implantation and PLM (data not shown). We believe that the role of Mn in stabilizing O in the Te sublattice is suppressed because of the rapid epitaxial regrowth rate of the PLM process estimated to be on the order of meters per second.

Fig. 2 shows a series of PR spectra from the 3.3% $O^+$-implanted $Zn_{0.88}Mn_{0.12}Te$ samples after PLM with fluence = 0.15 J/cm$^2$ followed by RTA for 10s at temperatures between 300 and 700ºC. A reduction in the energy shifts of both $E_-$ and $E_+$ can be observed at RTA temperature higher than 350ºC. This indicates that the $Zn_{0.88}Mn_{0.12}O_xTe_{1-x}$ alloys formed by implantation and PLM are thermally stable up to ~350ºC. At an RTA temperature of 700ºC, only the original $E_M$ transition is observed. This may suggest that most of the implanted O atoms that resided in the Te sites after PLM diffused out of the Te sites, possibly forming O bubbles. It is also worth noting that the BAC model predicts that as the $E_-$ transition approaches the localized O level, as in the case of the samples after RTA at temperatures between 400 and 555ºC, the nature of the lowest subband minimum becomes more localized-like. This can account for the observed broadening of the transition in Fig. 2.

The energy positions of $E_-$ and $E_+$ for the $Zn_{0.88}Mn_{0.12}O_xTe_{1-x}$ alloys with different x are plotted in Fig. 3. Data taken from samples implanted with different amount of O (1.65, 2.2 and 4.4%) as well as PLM with different energy fluences are also plotted on

Fig. 3. We note here that x decreases with increasing energy fluence higher than the melt threshold (~0.08 J/cm$^2$); possibly due to the longer melt duration and/or dilution through the deeper melt depth. The energy positions of the two transitions as predicted by the BAC model are also plotted as solid lines. Here, since the values of x were calculated from the $E_-$ transition no error bars are given for $E_-$. Given the broad linewidths of the $E_+$ transitions, they agree reasonably well with the calculated values for samples with various O mole fractions.

The new diluted II-VI oxide has technological potentials for photovoltaic applications. Efforts to improve the efficiency of solar cells have led to extensive experimental and theoretical studies of new materials and cell designs. To date, the highest power conversion efficiency of ~ 33% have been achieved with multijunction solar cells based on standard semiconductor materials [16-18]. It was recognized over thirty years ago that the introduction of states in a semiconductor band gap presents an alternative to multijunction designs for improving the power conversion efficiency of solar cells [19-21]. It was argued that deep impurity or defect states could play the role of the intermediate states for this purpose. Detailed theoretical calculations indicated that a single junction cell with a properly located band of intermediate states could achieve power conversion efficiencies up to 62% [20]. Efficiencies of up to 71.7% were predicted for materials with two bands of intermediate states [21]. However, difficulties in controlling the incorporation of high concentrations of impurity or defect states have thwarted prior efforts to realize such materials.

With multiple band gaps that fall within the solar energy spectrum, $Zn_{1-y}Mn_yO_xTe_{1-x}$ provides a unique opportunity for the assessment of the proposed multiband solar cell. The energy band structure and the density of states for the case of $Zn_{0.88}Mn_{0.12}O_xTe_{1-x}$ alloy (with x~0.01) are shown in Fig 4. A narrow band, $E_-$ of O derived extended states is separated from the conduction band by about 0.7 eV. Three types of optical transitions are possible in this band structure; (1) the transitions from the valence band to the $E_+$ subband with the absorption edge at $E_{V+}=E_+(k=0)-E_V(k=0)=2.56$ eV, (2) transitions from the valence band to $E_-$ subband with the edge at $E_{V-}=E_-(k=0)-E_V(k=0)=1.83$ eV and (3) the low energy transitions from $E_-$ to $E_+$ with the absorption edge at $E_{+-}=E_+(k=0)-E_-(k=0)=0.73$ eV. These three absorption edges span much of the solar spectrum, thus these alloys are good candidates for the multi-band semiconductors envisioned for high efficiency photovoltaic devices.

To further evaluate the suitability of the $Zn_{0.88}Mn_{0.12}O_xTe_{1-x}$ HMAs for solar cell applications we have calculated the solar cell power conversion efficiency for the material with the electronic band structure shown in Fig. 4. The calculations are based on the implementation of the detailed balance theory [22] to the case of a three-band semiconductor [20,21]. Even for this non-optimal band gap configuration we calculate a power conversion efficiency of 45%, which is higher than the ideal efficiency of any solar cell based on a single junction in a single-gap semiconductor and is comparable to the efficiency of triple-junction cells.

The great advantage of our material system is the alloy composition provides a means for varying the three optical transition energies in order to optimize solar cell performance. Our calculation shows that by increasing x in $Zn_{0.88}Mn_{0.12}O_xTe_{1-x}$ to slightly above 0.02 would increase the gap between $E_+$ and $E_-$ to 1 eV and lead to an optimum power conversion efficiency of 56%. Also, it is noted that changing the Mn content or replacing Mn with Mg may provide additional ways to vary the band structure for further optimization of solar cell performance.

In conclusion we have synthesized $Zn_{1-y}Mn_yO_xTe_{1-x}$ ternary (y=0) and quaternary (y=0.12) alloys by $O^+$-implantation followed by pulsed laser melting. Alloys with substitutional O mole fraction in the Te sublattice as high as 0.012 have been achieved. Optical transitions corresponding to the both the lower ($E_-$) and upper ($E_+$) conduction subbands resulting from the anticrossing interaction between the localized O states and the extended conduction states of the matrix are clearly observed in these diluted II-VI oxides. We demonstrate that these alloys fulfill the criteria for three band semiconductor that has been proposed as a means of making high efficiency, single-junction solar cells.

This work was supported by the Director, Office of Science, Office of Basic Energy Sciences, Division of Materials Sciences and Engineering, of the U. S. Department of Energy under Contract No. DE-AC03-76SF00098. MAS acknowledges support from an NSF Graduate Research Fellowship.

FIGURE CAPTONS

Fig. 1  Photomodulated reflectance (PR) spectra obtained from a series of 3.3% $O^+$-implanted $Zn_{0.88}Mn_{0.12}Te$ samples followed by pulsed laser melting with increasing energy fluence from 0.04-0.3 J/cm$^2$. The PR spectrum from an as-grown $Zn_{0.88}Mn_{0.12}Te$ crystal is also shown for comparison.

Fig. 2  PR spectra from $Zn_{0.88}Mn_{0.12}Te$ samples implanted with 3.3 % O followed by PLM with 0.15J/cm$^2$ and rapid thermal annealing (RTA) for 10sec at annealing temperatures from 300-700ºC.

Fig. 3  The energy positions of $E_-$ and $E_+$ for the $Zn_{0.88}Mn_{0.12}O_xTe_{1-x}$ alloys plotted against the O mole fractions x. The values of $E_-$ and $E_+$ calculated according to the band anticrossing model are plotted as solid lines.

Fig. 4  The calculated energy band structure (left panel) and density of states (right panel) for $Zn_{0.88}Mn_{0.12}O_xTe_{1-x}$ with x~0.01. The three possible optical transitions are indicated in the right panel.

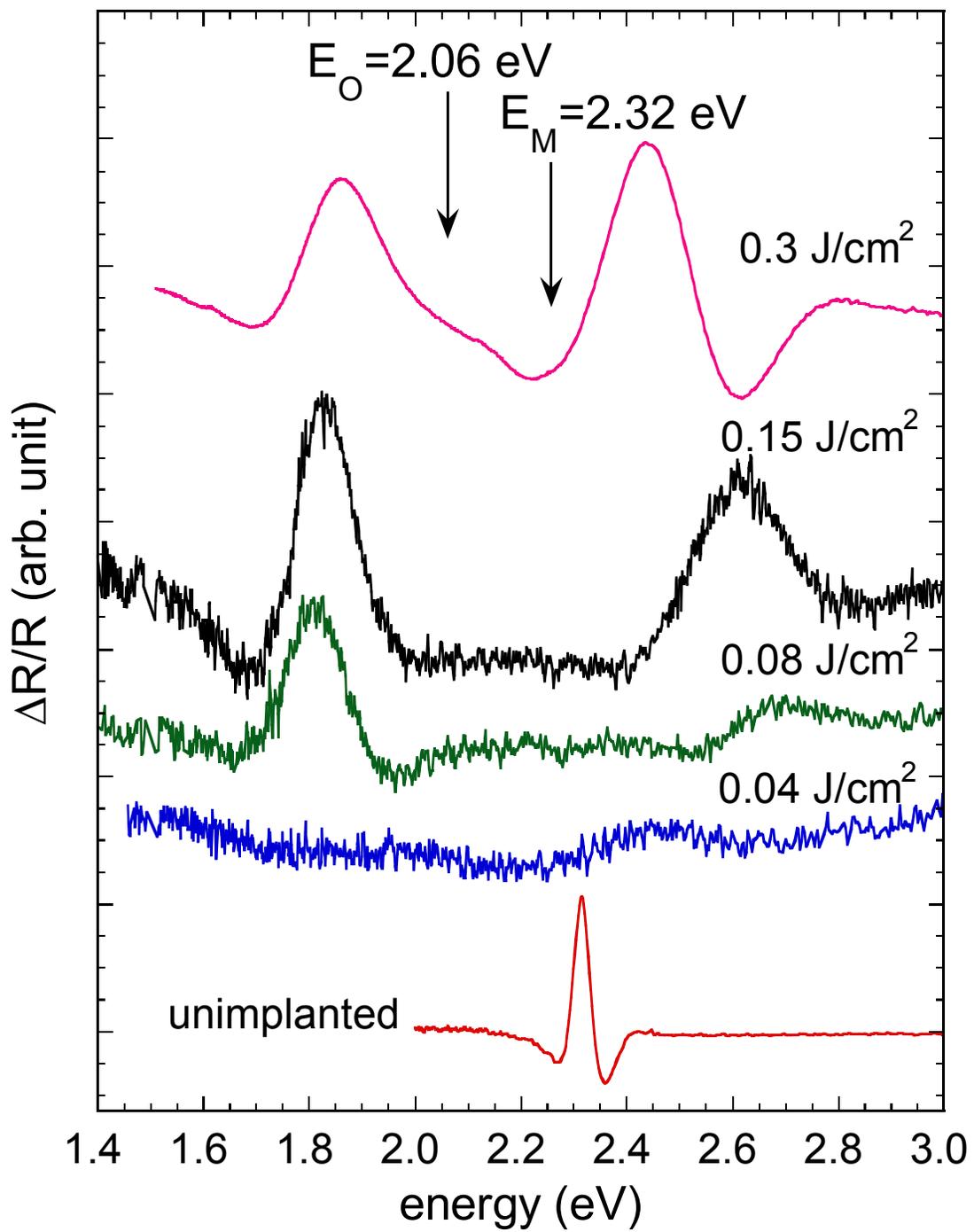

Figure 1

14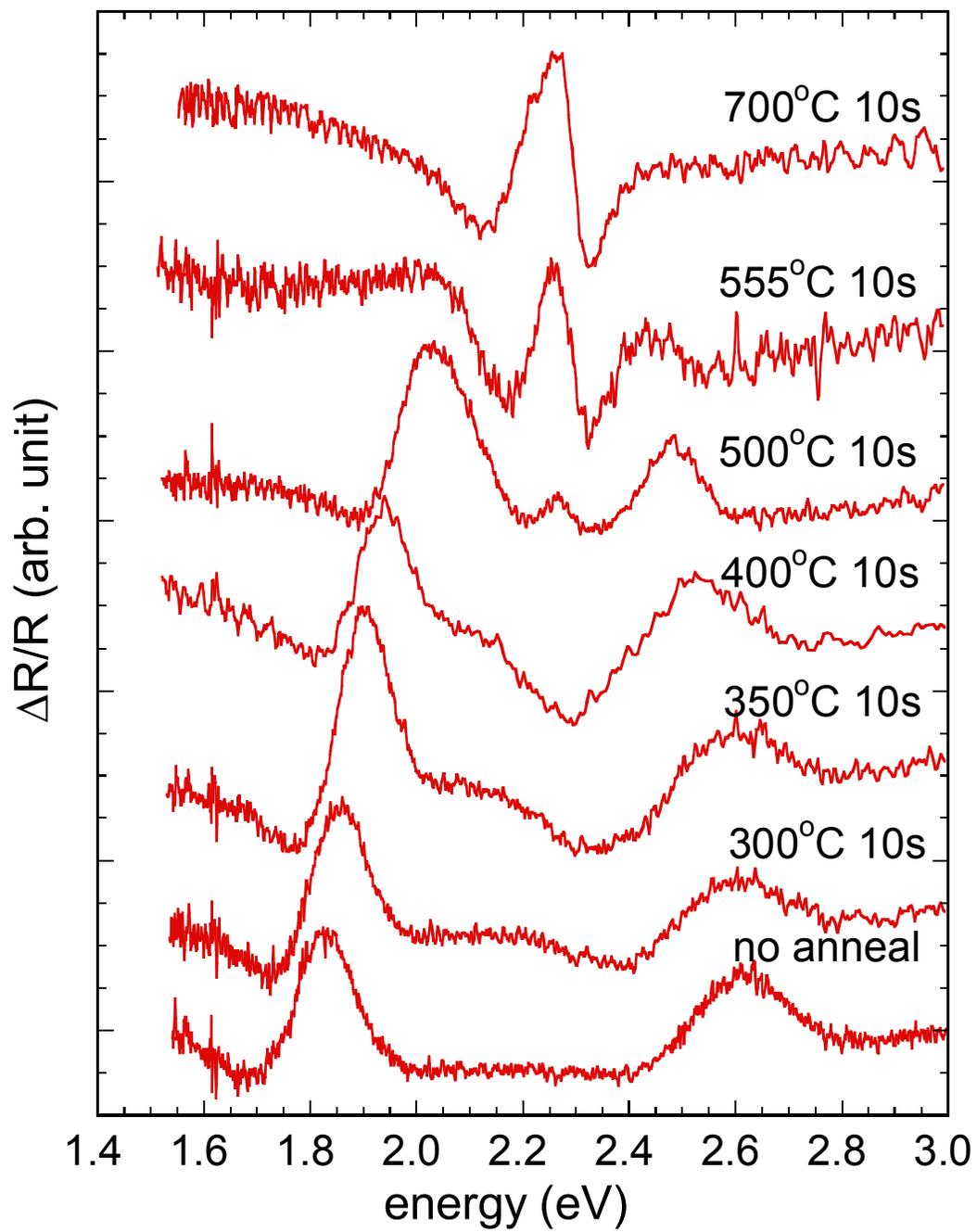

Figure 2

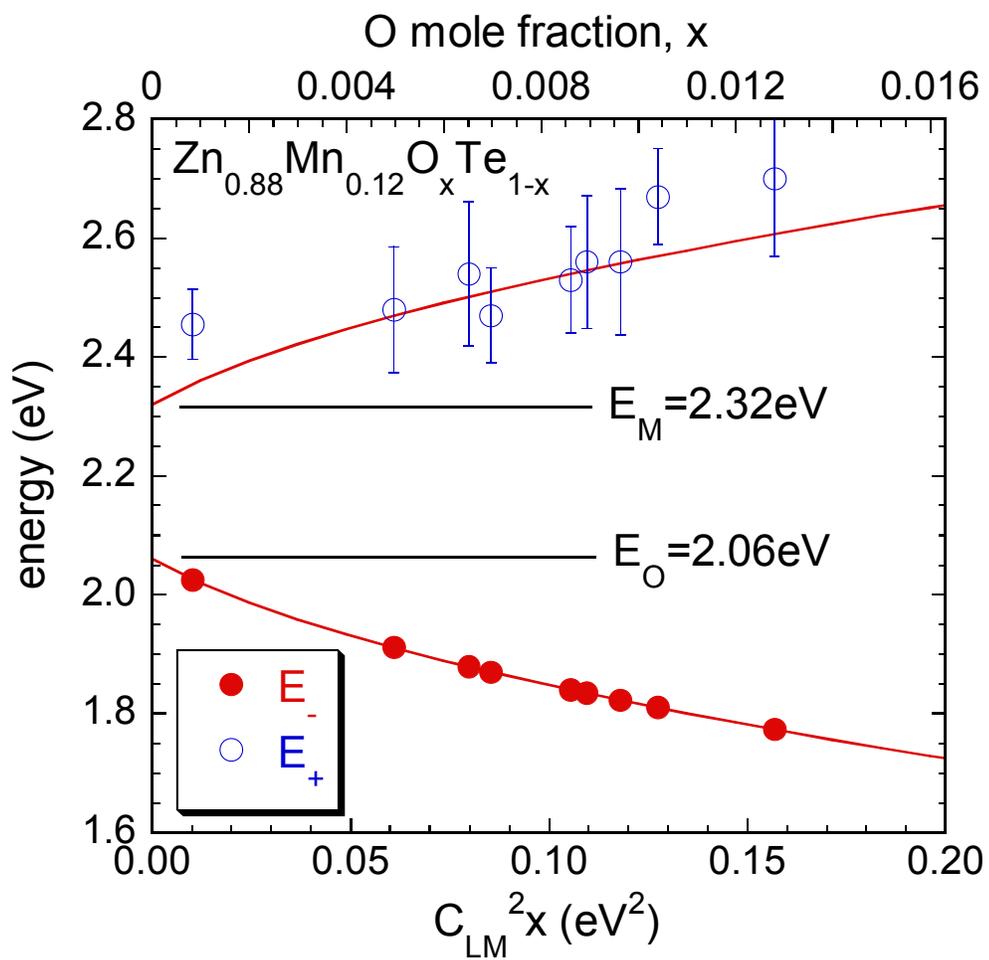

Figure 3

416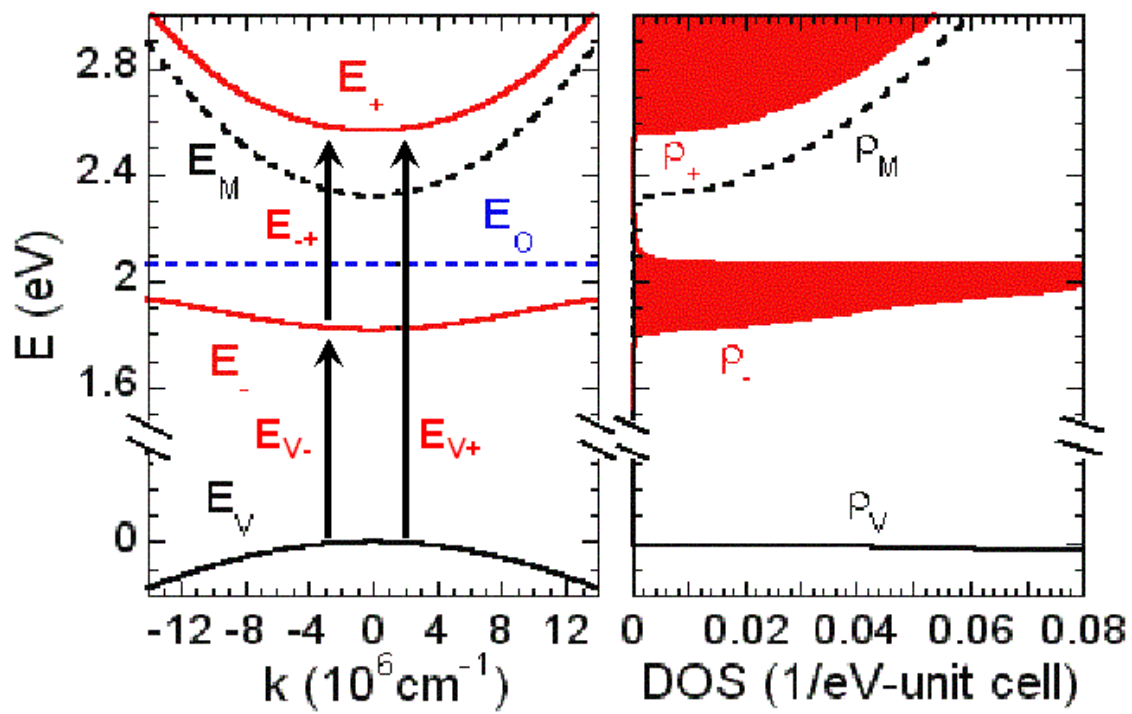

Fig. 4